\documentclass[11pt]{article}
\usepackage{epsfig}
\usepackage{amsfonts}
\usepackage{amsmath}
\usepackage{amssymb}
\usepackage{dsfont}
\usepackage{hyperref}
\usepackage{color}
\usepackage{cite}

\newcommand{\ep}{\epsilon}

\newcommand{\al}{\alpha}
\newcommand{\bt}{\beta}
\newcommand{\g}{\gamma}

\newcommand{\tb}{\bar \theta}

\newcommand{\Or}{\mathcal O}

\newcommand{\vL}{\ensuremath{\mathcal{L}}}
\newcommand{\vp}{\varphi}
\newcommand{\sq}{^{2}}

\newcommand{\dslash}[1]{#1 \llap{/\kern-0.5pt}}
\newcommand{\Dslash}[1]{#1 \llap{/\kern+1.5pt}}
\newcommand{\DDslash}[1]{#1 \llap{/\kern+2.3pt}}
\newcommand{\dslashh}[1]{#1 \llap{/\kern+1pt}}

\newcommand{\boldtau}{\mbox{\boldmath $\tau$}}
\newcommand{\boldpi}{\mbox{\boldmath $\pi$}}

\newcommand{\Ex}[1]{\cdot 10^{#1}}
\newcommand{\bea}{\begin{eqnarray}}
\newcommand{\eea}{\end{eqnarray}}
\newcommand{\bma}{\begin{pmatrix}}
\newcommand{\ema}{\end{pmatrix}}
\newcommand{\nn}{\nonumber}

\begin{document}
\begin{titlepage}

\begin{flushright}
LA-UR-16-28961\\
NIKHEF 2016-056
\end{flushright}

\vspace{2.0cm}

\begin{center}
{\LARGE  \bf 
An $\ep ' $ improvement from right-handed currents
}
\vspace{2.4cm}

{\large \bf  V. Cirigliano$^a$, W. Dekens$^{a,b}$, J. de Vries$^c$, and E. Mereghetti$^a$ } 
\vspace{0.5cm}

\vspace{0.25cm}

{\large 
$^a$ 
{\it Theoretical Division, Los Alamos National Laboratory,
Los Alamos, NM 87545, USA}}

\vspace{0.25cm}
{\large 
$^b$ 
{\it 
New Mexico Consortium, Los Alamos Research Park, Los Alamos, NM 87544, USA
}}

\vspace{0.25cm}
{\large 
$^c$ 
{\it 
Nikhef, Theory Group, Science Park 105, 1098 XG, Amsterdam, The Netherlands
}}

\end{center}

\vspace{1.5cm}

\begin{abstract}
Recent lattice QCD calculations of direct CP violation in $K_L \to \pi \pi$ decays indicate tension 
with the experimental results. Assuming this tension to be real, we investigate a possible beyond-the-Standard Model 
explanation via right-handed charged currents. By using chiral perturbation theory in combination with lattice QCD results, we 
accurately calculate the modification of $\ep'/\ep$ induced by right-handed charged currents and 
 extract values of the couplings that are necessary to explain the discrepancy, pointing to a scale around $10^2$ TeV.
We find that couplings of this size are not in conflict with constraints from other precision experiments, 
but next-generation hadronic 
electric dipole moment searches (such as neutron and ${}^{225}$Ra)  can falsify this scenario. We work out in detail a direct link, based on chiral perturbation theory, between CP violation in the kaon sector and electric dipole moments induced by right-handed currents which can be used in future analyses of left-right symmetric models.

\end{abstract}

\vfill
\end{titlepage}

\section{Introduction}

At the turn of the century the KTeV~\cite{AlaviHarati:2002ye,Abouzaid:2010ny}  and NA48~\cite{Batley:2002gn} collaborations
reported rather precise measurements of  $\ep'/\ep$    --  quantifying   direct CP violation in $K_L \to \pi \pi$ decays 
relative to CP violation in $K^0$-$\bar{K}^0$ mixing~\cite{Batley:2002gn,AlaviHarati:2002ye,Abouzaid:2010ny} -- 
with  world average  $(\ep' / \ep)_{\rm exp} =  (16.6 \pm 2.3) \times 10^{-4}$. 
A precise Standard Model (SM) prediction for $\ep'/\ep$ is a formidable task, requiring perturbative input on the Wilson coefficients 
in the weak Hamiltonian~\cite{Ciuchini:1992tj,Buras:1993dy,Buchalla:1995vs,Ciuchini:1995cd} 
and non-perturbative calculations of the relevant matrix elements 
(see  Refs.~\cite{Bertolini:1998vd,Buras:2003zz,Pich:2004ee,Cirigliano:2011ny} and references therein).

Thanks to advances in lattice QCD (LQCD) calculations of the hadronic matrix elements,  the long-known experimental result can be confronted  more and more confidently with the  SM predictions.  The  state-of-the-art  analysis of Ref.~\cite{Bai:2015nea} leads to  $(\ep' / \ep)_{\rm SM} =  (1.4 \pm 6.9) \times 10^{-4}$.
Taken at face value,  this result suggests a $2\sigma$ discrepancy between the SM prediction and the  observed value of $\ep'/\ep$.  While this  is in qualitative agreement with the finding of  Ref.~\cite{Buras:2015yba,Buras:2016fys}  
($(\ep' / \ep)_{\rm SM} =  (1.8 \pm 4.5) \times 10^{-4}$)  and of Ref.~\cite{Kitahara:2016nld}
($(\ep' / \ep)_{\rm SM} =  (1 \pm 5) \times 10^{-4}$)  
one should keep in mind that several analytic approaches to  $\ep'/\ep$ find results   consistent with the measurements  \cite{Bertolini:2000dy,Bijnens:2000im,Pallante:2001he}.

Assuming that the lattice result survives upcoming improved evaluations of the matrix elements  that will address all the lattice systematics,  it is  interesting and timely to investigate (i)  possible origins of the $\ep' / \ep$ enhancement in  beyond-the-Standard-Model (BSM) 
scenarios;  (ii) within such scenarios what correlations might emerge between $\ep' /\ep$ and other CP-violating (CPV) and flavor  observables. 
While the topic of BSM contributions to $\ep'/\ep$ has a long history 
(see for example~\cite{Gabbiani:1996hi,Buras:2000qz,Masiero:1999ub,Babu:1999xf,Barbieri:1999ax,Buras:1999da,Kagan:1999iq,Bertolini:2012pu,Bertolini:2014sua}),  
it has attracted renewed attention in the recent literature 
within $331$ models, non-standard $Z$ and $Z'$, as well as supersymmetric models~\cite{Buras:2015jaq,Buras:2015kwd,Tanimoto:2016yfy,Kitahara:2016otd,Endo:2016aws,Bobeth:2016llm}. 
Here  we discuss the possibility that the enhancement in $\ep' / \ep$ originates from  
right-handed charged-current (CC) interactions,  
parameterized by a single  gauge-invariant dimension-six operator,      
and study the  correlation with hadronic and atomic electric dipole moments (EDMs).  
While both $\ep' / \ep$~\cite{Bertolini:2012pu,Bertolini:2014sua}    and EDMs
\cite{Zhang:2007da,An:2009zh,Xu:2009nt,Maiezza:2014ala} 
have been studied in the context of left-right symmetric models~\cite{Mohapatra:1974hk,Senjanovic:1975rk},
 which induce  the right-handed CC operator of interest here, 
as far as we know the  enhancement of  $\ep' / \ep$ and its  correlation with  EDMs has  not been discussed  in the recent literature. 

We consider a setup in which right-handed CC interactions manifest  themselves  at low-energy 
through a single $SU(3) \times SU(2)\times U(1)$-invariant  dimension-6 operator~\cite{Buchmuller:1982ye,Grzadkowski:2010es}, namely
\begin{equation}
\vL_{\rm eff}  = {\cal L}_{\rm SM} +  \frac{2}{v^2} i \tilde{\vp}^{\dagger} D_{\mu} \vp \, \bar{u}^i_R \gamma^\mu \,\xi_{i j} d^j_R 
+  \mathrm{h.c.}, 
  \ \to \  {\cal L}_{\rm SM} + \frac{g}{\sqrt 2}\bigg[\xi_{ij}\,\bar u^i_R \g^\mu  d^j_R\,W_\mu^+  \bigg]\left(1+\frac{h}{v}\right)^2 +\text{h.c.},
\ \ \ 
\label{dim6edms}
\end{equation}
where $D_\mu$ is the covariant derivative, $\vp$ is the Higgs doublet, $i$ and $j$ are generation indices, 
and  $g$ is the $SU(2)$ gauge coupling. 
The second form in the above equation  is obtained after electroweak symmetry breaking in the unitary gauge. 
The matrix $\xi_{ij}$
(not necessarily unitary)  scales as $\xi_{ij} \sim \mathcal O(v^2/\Lambda^2)$, where $\Lambda$ is the scale of new physics. 
Note that this operator arises as the single dominant low-energy manifestation of CP violation within $P$-symmetric left-right symmetric models\footnote{While this is correct for EDMs, left-right models also generate operators of the form $(\bar s_R \g_\mu u_R)\, (\bar u_R \g^\mu d_R)$ that contribute to $\ep'$. However, the matrix elements of these operators are chirally suppressed, such that their contributions can be neglected with respect to those from the operator under consideration here \cite{Bertolini:2012pu,Bertolini:2013noa}.}, 
but we study its phenomenological impact without reference to any underlying model. 

In our analysis we focus on the couplings  $\xi_{ud}$ and $\xi_{us}$, assuming that they have complex phases.  
At the weak scale we integrate out the $W$ boson and discuss the implications of the  resulting  CPV  four-quark operators, 
both with $\Delta S=1$ (contributing to $\ep'/\ep$) and with $\Delta S = 0$ (contributing to hadronic EDMs). 
Since the induced four-quark operators  belong to the irreducible $(8_L, 8_R)$ representation of the chiral group $SU(3)_L \times SU(3)_R$, 
a number of implications emerges: 

\begin{itemize}
\item One is able to  relate the  matrix elements required to evaluate the new  contribution to 
$\ep^\prime/\ep$ to the matrix elements of the electroweak penguin operators $\mathcal Q_{7,8}$~\cite{Chen:2008kt,Bertolini:2013noa}.
Here we update previous analyses using the recent LQCD input~\cite{Blum:2012uk,Blum:2015ywa}. 
\item  At leading  order in chiral perturbation theory (ChPT), we are able to relate the $K \rightarrow \pi \pi$   matrix elements  to CPV meson-baryon couplings, which  provide a leading contribution to hadronic and nuclear EDMs.  Within $\mathcal O(1)$ hadronic uncertainties, this enforces a correlation between  
$\ep'/\ep$  and EDMs,   which we explore phenomenologically.  
Anticipating the main results, we find that for ranges of  $\xi_{ud}$ and $\xi_{us}$ that  lead to the desired  $\ep'/\ep$ enhancement,  
EDMs of the neutron, deuteron, $^{129}$Xe,  $^{199}$Hg,  and $^{225}$Ra are predicted 
within reach of next generation searches and with a definite  pattern.
\end{itemize}

In Sec.~\ref{LowEnergy} we derive the low-energy interactions resulting from the right-handed current operator in Eq.~\eqref{dim6edms}, 
both at the quark level and hadronic level in ChPT.  
We subsequently discuss the contributions to $\ep' / \ep$ and  $\ep_K$ in Sec.~\ref{sec:eps'},   the contributions to hadronic / nuclear EDMs in Sec.~\ref{sec:EDMs}, 
and the resulting phenomenology in  Sec.~\ref{sec:analysis}, before concluding in Sec.~\ref{sec:conclusion}.

\section{Low-energy Lagrangian induced by right-handed  currents}\label{LowEnergy}

At the weak scale,  after integrating out the $W$ boson at tree-level, the effective Lagrangian in Eq.~\eqref{dim6edms} induces 
both semi-leptonic and four-quark operators, 
\begin{equation}
\label{eq:Leff2}
\tilde{\cal L}_{\rm eff}  =    \tilde{\cal L}_{\rm eff,SM} 
-\frac{4 G_F}{\sqrt{2}} \,  \left(  \xi^*_{ij}\,  \bar d^j \gamma^\mu P_R u^i \, \bar \nu \gamma_\mu P_L l  + \textrm{h.c.} \right) 
 - \sum^{2}_{a=1} \,\left(  C^{ij\, lm}_{a\, LR} \mathcal O^{ij\, lm}_{a\, LR} + C^{{ij\, lm}\,*}_{a\, LR} \big(\mathcal O^{ij\, lm}_{a\, LR}    \big)^\dagger  \right)\ ,      
\end{equation}
where $\tilde{\cal L}_{\rm eff, SM}$ is the SM effective Lagrangian below the weak scale, 
$P_{L,R} = (1\mp\gamma_5)/2$, $i$-$m$ are flavor indices, and the four-quark operators are defined as
\begin{eqnarray}\label{eq:4q1}
\mathcal O^{ij\, lm}_{1\, LR} = \bar d^m \gamma^\mu P_L u^l \, \bar u^i \gamma_\mu P_R d^j\ , \qquad  \mathcal O^{ij\, lm}_{2\, LR}  =\bar d_\al^m   \gamma^\mu P_L u_\bt^l \, \bar u_\bt^i  \gamma_\mu P_R d_\al^j\ ,
\end{eqnarray}
where  $\al,\,\bt$ are color indices. 
Tree-level  matching at the $W$ boson mass scale gives
\begin{eqnarray}\label{eq:4q2}
C^{ij\, lm}_{1\, LR}(m_W)  =\frac{4 G_F}{\sqrt{2}}  V^*_{lm} \xi_{i j}\ , \qquad C^{ij\, lm}_{2\, LR}(m_W) = 0 \ .
\end{eqnarray}
The couplings of the  four-fermion operators in $\tilde{\cal L}_{\rm eff,SM}$   scale as two inverse powers of the electroweak scale, $\sim 1/v\sq$, while the `left-right' operators induced by the right-handed currents scale as two inverse powers of the scale of new physics, $C_{i\, LR}\sim \xi/v\sq\sim 1/\Lambda\sq$. We neglect operators that are quadratic in $\xi$ and are suppressed by $v^2/\Lambda\sq$ with respect to the linear terms. 

As evident from Eq.~\eqref{eq:Leff2},  the leading low-energy implications of the new couplings $\xi_{ij}$ are expected in semi-leptonic transitions and non-leptonic 
transitions with both $\Delta F =0$ and $\Delta F = 1$  ($\Delta F = 2$ four-quark operators arise at loop level and will be briefly discussed in Sec.~\ref{sec:eps'}).  
Here we focus on CPV effects and note that 
the operators $\mathcal O_{i\, LR}$  lead to CP violation even if all generation indices are the same. Therefore the operators with $\Delta F =0$ that contain only the light quarks 
can generate hadronic  and nuclear EDMs. 
Four such operators exist
\begin{eqnarray}\label{eq:4q4}
\mathcal L_{\textrm{EDM}} &=& - i \Big(   \textrm{Im}\, C^{ud\, ud}_{1\, LR} \, \bar d \gamma^\mu P_L  u \, \bar u \gamma_\mu P_R d   
+ \textrm{Im}\, C^{ud\, ud}_{2\, LR} \,  \bar d_\al \gamma^\mu P_L  u_\bt \, \bar u_\bt \gamma_\mu P_R d_\al  \nn
 \\ & & +  \textrm{Im} \, C^{us\, us}_{1\, LR}  \,  \bar s \gamma^\mu P_L  u \, \bar u \gamma_\mu P_R s 
 +  \textrm{Im} \, C^{us\, us}_{2\, LR}  \, \bar s_\al \gamma^\mu P_L  u_\bt \, \bar u_\bt \gamma_\mu P_R s_\al - \text{h.c.} \Big) \ .
\end{eqnarray}
On the other hand,  CPV  $\Delta S =1$ operators contributing to  $\ep^\prime/\ep$ are 
\begin{eqnarray}\label{eq:4q5}
\mathcal L_{\ep^\prime} &=& - i  \Big( 
   \textrm{Im}\, C_{1\, LR}^{us\, ud} \, \bar d \gamma^\mu P_L u \, \bar u \gamma_\mu P_R s   +  \textrm{Im}\, C_{2\, LR}^{us\, ud} \,  \bar d_\al\gamma^\mu P_L u_\bt \, \bar u_\bt\gamma_\mu P_R s_\al   \nn \\
&&+  \textrm{Im}\, C_{1\, LR}^{ud\, us} \, \bar s \gamma^\mu P_L u \, \bar u \gamma_\mu P_R d
+  \textrm{Im}\, C_{2\, LR}^{ud\, us} \,  \bar s_\al\gamma^\mu P_L u_\bt \, \bar u_\bt \gamma_\mu P_R d_\al 
-{\rm h.c.}\Big)\ .
\end{eqnarray}
While the semi-leptonic operators in Eq.~\eqref{eq:Leff2}   do not undergo QCD renormalization group evolution, 
evolving  the LR operators to the hadronic scale \cite{Cho:1993zb,An:2009zh,Hisano:2012cc,Dekens:2013zca} leads to 
\bea
C_{1\, LR}^{ijlm}(3\, {\rm GeV})= 0.9\, C_{1\, LR}^{ijlm}(m_W)\ ,\qquad
C_{2\, LR}^{ijlm}(3\, {\rm GeV})=  0.4\,C_{1\, LR}^{ijlm}(m_W)+1.9\,C_{2\, LR}^{ijlm}(m_W)\ .
\eea

At the scale of a few GeV the effective Lagrangian is given by 
\begin{equation}\label{LagUnaligned}
\mathcal L = \mathcal L^{\rm QCD}_{m_q=0} - e^{i \rho} \bar q_L \mathcal M q_R - e^{-i \rho} \bar q_R \mathcal M q_L 
 - \theta \frac{g_s^2}{64 \pi^2} \varepsilon^{\mu \nu \alpha \beta} G^a_{\mu\nu} G^a_{\alpha \beta} + \tilde{\cal L}_{\rm eff,SM} + \mathcal L_{EDM}+\vL_{\ep'}\, ,
\end{equation}
where $q$ is a triplet of quark fields $q = (u, d, s)$,   $G^a_{\mu \nu}$ is the gluon field strength, $g_s$ is the strong coupling constant, $\mathcal M = {\rm diag}(m_u,\,m_d,\, m_s)$, and we include a common phase $\rho$ and the QCD $\theta$ term as they will play a role later. 
The Lagrangian in Eq.\ \eqref{LagUnaligned} includes operators that explicitly break chiral, isospin, and time-reversal symmetry and it therefore induces vacuum misalignment \cite{
Dashen:1970et,deVries:2012ab}, which manifests in the coupling of  the neutral mesons $\pi^0, K^0, \eta$  to the vacuum.
To avoid this, we perform an anomalous axial $U(1)_A$ rotation to eliminate the gluonic theta term, and a subsequent non-anomalous $SU(3)$ axial rotation to  eliminate leading-order tadpoles. The resulting Lagrangian can be cast in the following form
\begin{equation}
\mathcal L = \mathcal L^{\rm QCD}_{m_q = 0} - \bar q \mathcal M q + \bar q \left[ m_{*} (\bar \theta - \bar \theta_{\rm ind})+ m_3 t_3 + m_6 t_6 + m_8 t_8\right]i \gamma_5  q + \tilde{\cal L}_{\rm eff,SM} + \mathcal L_{EDM}+\vL_{\ep'}\ ,
\end{equation}
where we introduced the reduced quark mass $m_* = (1/m_u + 1/m_d + 1/m_s)^{-1}$, $\bar \theta = \theta -3 \rho$, the physical combination of phases in the SM, and four BSM quantities,  $\bar{\theta}_{\rm ind}$ and  $m_{3,6,8}$, that need to be determined by vacuum alignment.

The effective chiral Lagrangian associated to ${\cal L}^{\rm QCD}$ and $\tilde{\cal L}_{\rm eff,SM}$ is well known. 
At leading chiral order   $\tilde{\cal L}_{\rm eff,SM}$ induces two $\Delta S =1$ operators, belonging to the $(8_L,1_R)$ and $(27_L,1_R)$  
representations of $SU(3)_L \times SU(3)_R$, and we explicitly give below only the dominant $(8_L,1_R)$  operator. 
To construct the chiral Lagrangian induced by $\mathcal L_{EDM}$ and $\vL_{\ep'/\ep}$, we note that the four-quark operators in Eqs.\ \eqref{eq:4q4} and \eqref{eq:4q5} 
can schematically be written as $(\bar q \g_\mu t^a P_L q)(\bar q \g^\mu t^b P_R q)$,  with $t^{a,b}$  the generators of $SU(3)_{L,R}$, and 
belong to the same irreducible $(8_L,\, 8_R)$ representation of  $SU(3)_L\times SU(3)_R$. 
These left-right operators become formally invariant under a chiral transformation ($L, R \in SU(3)_{L,R}$) , $q_L \rightarrow L q_L$ and  $q_R \rightarrow R q_R$, by assigning the transformation properties $t^a \rightarrow L t^a L^\dagger$, $t^b \rightarrow R t^b R^\dagger$. 
To leading order,  the resulting mesonic Lagrangian is
\begin{eqnarray}\label{mesonEps'}
& & \mathcal L_\pi = \frac{F_0^2}{4} \textrm{Tr} \left(\partial_\mu U \partial^\mu U^\dagger  \right) +\frac{F_0^2}{4} \textrm{Tr} \left( U \chi^\dagger + U^\dagger \chi \right) 
+ F_0^4  \left( G_8 \textrm{Tr}\left( (t_6 - i t_7) \partial_\mu U^\dagger \, \partial^\mu U \right) +{\rm h.c.} \right)
\nn\\
&& + \frac{F_0^4}{4}   \textrm{Tr}\left( U^\dagger t^b Ut^a  \right)\sum_{i=1,2} \mathcal A_{i\, LR}\bigg[
C_{i\, LR}^{ud\, ud} (\delta_{a1}-i\delta_{a2})(\delta_{b1}+i\delta_{b2})+C_{i\, LR}^{us\, us}(\delta_{a4}-i\delta_{a5})(\delta_{b4}+i\delta_{b5})\nn\\
&&+C_{i\, LR}^{ud\, us}(\delta_{a4}-i\delta_{a5})(\delta_{b1}+i\delta_{b2})+C_{i\, LR}^{us\, ud}(\delta_{a1}-i\delta_{a2})(\delta_{b4}+i\delta_{b5})+{\rm h.c.}\bigg]\ ,
\end{eqnarray}
where  $U$ is the usual matrix of the pseudo-Nambu-Goldstone (pNG) boson fields
\begin{equation}\label{eq:2.3}
U = u(\pi)^2  = \exp\left(\frac{2 i \pi}{F_0}\right), \qquad
\pi =\frac{1}{\sqrt{2}} \left( \begin{array}{c c c} 
\frac{\pi_3}{\sqrt{2}} + \frac{\pi_8}{\sqrt{6}}  & \pi^+ 						& K^+ \\
\pi^-						& - \frac{\pi_3}{\sqrt{2}} + \frac{\pi_8}{\sqrt{6}} 	& K^0 \\
K^-						& \bar{K}^0					& - \frac{2}{\sqrt{6}}\pi_8 
\end{array} \right)\, ,
\end{equation}
and 
\begin{equation}\label{chi}
\chi = 2 B \left( \mathcal M + i \left( m_* \left(\bar\theta - \bar\theta_{\textrm{ind}}\right) + m_3  t_3  +m_6t_6+ m_8 t_8  \right)\right)\, .
\end{equation}
From the $K \rightarrow \pi \pi$ amplitudes, one can  infer $|G_8| \simeq 0.8 \,G_F$~\cite{Cirigliano:2011ny}. 
$\mathcal A_{1\, LR}$ and $\mathcal A_{2\, LR}$ are low-energy constants related to the $\Or_{1\, LR}$ and $\Or_{2\, LR}$ operators, respectively, and $F_0$ is the pNG decay constant in the chiral limit (we use $F_{\pi} = 92.2$ MeV and $F_{K} = 113$ MeV \cite{Agashe:2014kda} for the physical decay constants). Finally, the vacuum is aligned when we choose $m_i$ and $\bar\theta_{\rm ind}$ as follows, 
\begin{eqnarray}\label{align}
\bar\theta_{\textrm{ind}} &=& 
-\sum_{i=1,2} r_i {\rm Im}\, \left(\frac{m_d-m_u}{2m_um_d}C_{i\, LR}^{ud\, ud}+\frac{m_s-m_u}{2m_um_s}C_{i\, LR}^{us\, us}\right),\nn\\
m_3 &=& - \sum_{i=1,2}r_i {\rm Im}\,\left( C_{i\, LR}^{ud\, ud} + \frac{1}{2} C_{i\, LR}^{us\, us} \right) ,\nonumber \\
m_6 &=&  \frac{1}{2}\sum_{i=1,2}r_{i}{\rm Im}\,\left(C_{i\, LR}^{ud\, us}+ C_{i\, LR}^{us\, ud}\right), \nn\\
m_8 &=& -\frac{\sqrt{3}}{2}\sum_{i=1,2}  r_i \textrm{Im} \, C_{i\, LR}^{us\, us}\ ,
\end{eqnarray}
where $r_{i}=\frac{F_0\sq}{B} \mathcal A_{i\, LR}$. 
We introduced $\bar \theta_{\rm ind}$ because $\bar \theta \rightarrow \bar \theta_{\rm ind}$ if the strong CP problem is resolved via the Peccei-Quinn mechanism~\cite{Peccei:1977hh}.

The Lagrangian  \eqref{mesonEps'} induces new contributions to EDMs and $\ep'/\ep$  originating from the  left-right  operators, 
controlled by the low-energy constants  $\mathcal A_{1,2\, LR}$, 
which can be related to the $K \to \pi \pi$ matrix elements 
of the electroweak penguin operators ${\mathcal Q}_{7,8}$~\cite{Bijnens:1983ye,Buchalla:1995vs} 
\bea
\mathcal Q_7 = 6 (\bar s \g^\mu P_L d) \sum_{q=u,d,s} Q_q (\bar q\g_\mu P_R q)\ ,
\qquad \mathcal Q_8 =  6 (\bar s_\al \g^\mu P_L d_\bt) \sum_{q=u,d,s} Q_q (\bar q_\bt\g_\mu P_R q_\al) \ , 
\eea
where $Q_q$ denotes the electric charge.
Both $\mathcal Q_{7}$ and $\mathcal Q_{8}$ can be written as combinations of $(\bar q \g_\mu t^a P_L q)(\bar q \g^\mu t^b P_R q)$ and therefore belong to the same irreducible representation as the left-right operators in Eqs.\ \eqref{eq:4q4} and \eqref{eq:4q5}~\cite{Chen:2008kt,Bertolini:2013noa}.
As a result, the mesonic Lagrangian they induce is similar to Eq.\ \eqref{mesonEps'}:  it involves different $SU(3)_{L,R}$ indices $a,\, b$, 
but it comes with the same low-energy constants, $\mathcal A_{1,2\, LR}$.
The contributions of $\mathcal Q_{7,8}$ to the $K\to \pi\pi$ amplitudes derived from this Lagrangian, together with the lattice results of Ref.\ \cite{Blum:2012uk}, then allow for the extraction of $\mathcal A_{1,2\, LR}$.  Working to leading order in ChPT  we obtain
\begin{eqnarray}
\mathcal A_{1\, LR}(3\, \textrm{GeV}) &=& \frac{1}{\sqrt{3} F_0}  \, \langle (\pi\pi)_{I=2}|\mathcal  Q_7|K^0\rangle
 + \mathcal O\left(m^2_K\right) \simeq (2.2\pm0.13) \, {\rm GeV}^2\ ,\nn\\
\mathcal A_{2\, LR}(3\, \textrm{GeV}) &=& \frac{1}{\sqrt{3} F_0}  \, \langle (\pi\pi)_{I=2}|\mathcal  Q_8|K^0\rangle
 + \mathcal O\left(m^2_K\right)\simeq (10.1\pm0.6) \, {\rm GeV}^2\ . 
\label{LEC}
\end{eqnarray}

\section{Contributions to $\ep' / \ep$ and $\ep_K$}\label{sec:eps'}

Direct CP violation in $K_L\to \pi\pi$ decays is quantified by $\ep'$, which can be expressed as 
\bea
{\rm Re}\, \bigg(\frac{\ep'}{\ep}\bigg) = {\rm Re}\,\bigg(\frac{i \omega e^{i(\delta_2-\delta_0)}}{\sqrt{2}\ep}\bigg)\bigg[\frac{{\rm Im}\,A_2}{{\rm Re}\,A_2}-\frac{{\rm Im}\,A_0}{{\rm Re}\,A_0}\bigg]\label{epsPrime}\ , 
\eea
where $A_{0,2} e^{i\delta_{0,2}}$
are the amplitudes for final-state pions with total isospin $I=0,2$, 
(strong phases are denoted by $\delta_{0,2}$) 
and $\omega \equiv { {\rm Re}\,A_2} / {{\rm Re}\,A_0}$.

In the SM, $A_{0}$ and $A_{2}$ are sensitive to contributions from CC operators, $\mathcal Q_{1-2}$, strong penguin operators, $\mathcal Q_{3-6}$, and electroweak penguin operators, $\mathcal Q_{7-10}$. The values of their next-to-leading-order (NLO) Wilson coefficients have been calculated in Refs.\ \cite{Ciuchini:1992tj,Buras:1993dy,Buchalla:1995vs,Ciuchini:1995cd}, while lattice determinations of the necessary matrix elements are given in Refs.\ \cite{Blum:2015ywa,Bai:2015nea,Blum:2012uk}. Combining these results with the following experimental values \cite{Agashe:2014kda}
${\rm Re}\, A_0 = 33.201\Ex{-8} \,{\rm GeV}$, 
${\rm Re}\, A_2 = 1.479\Ex{-8} \,{\rm GeV} $,  
$\omega = 0.04454$, 
$|\ep |= (2.228\pm0.011)\Ex{-3}$, 
${\rm Arg}\, \ep = 0.75957\, {\rm rad}$ 
and lattice determinations of the strong phases, $\delta_0 = (23.8\pm4.9\pm 1.2)^\circ$, $\delta_2 = -(11.6\pm2.5\pm 1.2)^\circ$, leads to the SM prediction \cite{Bai:2015nea}
\bea
{\rm Re}\, \bigg(\frac{\ep'}{\ep}\bigg)_{\rm SM}=(1.38\pm 5.15\pm 4.59)\Ex{-4} \simeq (1.4\pm 6.9)\Ex{-4}\,. 
\eea

The contributions of the right-handed currents to $\ep'/\ep$ can be calculated using  Eq.\ \eqref{mesonEps'}.
 Such a determination  in principle still  suffers from  higher-order, $\Or(m_K\sq)$, uncertainties.
Fortunately, after an isospin decomposition, the $I=3/2$ parts of the LR operators, $O_{1\, LR}^{ud\, us}$ and $O_{2\, LR}^{ud\, us}$,  coincide  with those of $\mathcal Q_{7}$ and $\mathcal Q_8$, respectively. Isospin symmetry therefore implies a stronger relation between the contributions of the left-right operators to the $I=2$ amplitude and the matrix elements of  $\mathcal Q_{7,8}$ \cite{Chen:2008kt,Bertolini:2013noa}, 
subject to percent-level   $\Or((m_d-m_u)/\Lambda_\chi)$  and   $\Or(\alpha/\pi)$  corrections. 
 The resulting expression for the $I=2$ amplitude is 
\bea
{\rm Im}\, A_2(\xi) =\frac{F_0}{2\sqrt{6}}{\rm Im}\,\bigg[\big(C_{1LR}^{udus}-C_{1LR}^{usud^*}\big)\mathcal A_{1\, LR}
+\big(C_{2LR}^{udus}-C_{2LR}^{usud^*}\big)\mathcal A_{2\, LR}\bigg]\ .\label{eq:A2expr}
\eea
For the $I=0$ amplitude to leading order in ChPT, Eq.\ \eqref{mesonEps'} predicts
\bea\label{A0A2relation}
A_0(\xi) = -2\sqrt{2}A_2(\xi)\ ,
\eea
which can be affected by $\mathcal O(m_K^2)$ corrections. We thus find 
\bea\label{finaleps'}
{\rm Re}\, \bigg(\frac{\ep'}{\ep}\bigg)={\rm Re}\, \bigg(\frac{\ep'}{\ep}\bigg)_{\rm SM}+{\rm Re}\,\bigg(\frac{i \omega e^{i(\delta_2-\delta_0)}}{\sqrt{2}\ep}\bigg)\bigg[\frac{{\rm Im}\,A_2(\xi)}{{\rm Re}\,A_2}-\frac{{\rm Im}\,A_0(\xi)}{{\rm Re}\,A_0}\bigg]\ ,
\eea
where we use the experimental values\footnote{
An $\mathcal O(1)$ positive shift to $\ep'/\ep$  could be explained by an $\mathcal O(1)$ correction to ${\rm Re}\,A_{0,2}$.  
Such large corrections to ${\rm Re} \, A_{0,2}$  are not plausible in this scenario as the real parts of the couplings $\xi_{ud,us}$ are 
constrained at the $10^{-3}$ level by semi-leptonic transitions~\cite{Buras:2010pz,RHC}.} for ${\rm Re}\,A_{0,2}$. 
The expression for $A_0(\xi)$ in Eq.\ \eqref{A0A2relation} might suffer from relatively large $SU(3)$ corrections. However,  
the dominant $\xi$ contribution to $\ep'$ arises from $A_2 (\xi)$, while the $A_0(\xi)$ term is suppressed by $2\sqrt{2}\omega\simeq 0.1$. We therefore expect Eq.~\eqref{finaleps'}, with  ${\rm Im}\, A_{0,2}(\xi)$ from Eqs.~\eqref{eq:A2expr}-\eqref{A0A2relation}, to be accurate up to the lattice uncertainties in Eq.~\eqref{LEC}.

The imaginary parts of  $\xi_{ud}$ and $\xi_{us}$ also induce corrections to CP violation in $K_0 -\bar{K}_0$ mixing.
The time evolution of the $(K_0, \bar{K}_0)$ system is governed by the Hamiltonian $H  = M - i \Gamma/2$, where $M$ and $\Gamma$ are 2 $\times$ 2 hermitian matrices.
Indirect CP violation arises from the weak phase difference between the off-diagonal elements of $M$ and $\Gamma$, and it is parameterized by $\ep_K$ \cite{Buchalla:1995vs}
\begin{equation}
\ep_K = \frac{e^{i \frac{\pi}{4}}}{\sqrt{2}} \left( \frac{\textrm{Im}\,M_{12}}{2 \textrm{Re}\, M_{12}} - \frac{\textrm{Im}\,\Gamma_{12}}{2 \textrm{Re}\, \Gamma_{12}}  \right)
\simeq \frac{e^{i \frac{\pi}{4}}}{\sqrt{2}} \left( \frac{\textrm{Im}\,M_{12}}{\Delta m_K} + \frac{\textrm{Im}\, A_0}{\textrm{Re}\, A_0}  \right),
\end{equation}
where in the second step we replaced $2 \textrm{Re}\, M_{12}$ with $\Delta m_K  = m_{K_L} - m_{K_S}$, and used the fact that the neutral kaon decay width is saturated   
by the decay into two pions in the isospin 0 channel.
The impact of right-handed CC on $\textrm{Im}\, A_0$ was discussed above.  We now focus on corrections to $\textrm{Im}\, M_{12}$, 
which originate at  both  long- and short-distance.

Long-distance corrections to $M_{12}$ arise from two insertions of $\Delta S = 1$ operators in the chiral Lagrangian \eqref{mesonEps'}.
Both the SM operator $G_8$ and the LR operators $\mathcal A_{i\, LR}$ induce mixing between the neutral kaons and $\pi_0$ and $\eta$. 
Therefore, the $K_0 \rightarrow \bar{K}_0$ amplitude receives a tree-level contribution from diagrams in which the $K_0$ mixes into a pion or a $\eta$ meson,
which then mixes into a $\bar{K}_0$.
For these diagrams, we find
\begin{equation}\label{eq:epsK0}
2 m_K \textrm{Im}\, M_{12}(\xi) 
= - \frac{1}{2} F_0^4 G_8 \sum_{i} \mathcal A_{i\, LR} \, \left(\textrm{Im}\,C_{i\, LR}^{ud\,us} - \textrm{Im}\,C_{i\, LR}^{us\,ud}\right)  \frac{m_{K^0}^2 (4 m_{K^0}^2 - 3 m_\eta^2 - m_{\pi^0}^2)}{(m_{K^0}^2 - m^2_\eta) \, (m^2_{K^0} - m^2_{\pi^0})},
\end{equation}
where we neglected the contribution from the imaginary part of $G_8$ and  real part of  $\xi_{ud,us}$.
Analogously to the SM case~\cite{Buras:2010pza},   
the Gell-Mann--Okubo formula implies that Eq. \eqref{eq:epsK0} vanishes at LO  and starts contributing at NLO.
At the same order in ChPT
 one has  to consider loop diagrams with two insertions of  three-pNG vertices from Eq. \eqref{mesonEps'} 
and local counterterms from subleading $\Delta S = 2$  operators in the chiral Lagrangian. 
They have the generic form 
\begin{equation}\label{eq:epsK0-NLO}
2 m_K \textrm{Im}\, M_{12}(\xi) 
\sim  \frac{1}{2} F_0^4 G_8 \sum_{i} \mathcal A_{i\, LR} \, \left(\textrm{Im}\,C_{i\, LR}^{ud\,us}  \pm  \textrm{Im}\,C_{i\, LR}^{us\,ud}\right)
\  \frac{m_{K^0}^2}{(4\pi F_0)^2}  \times   f_\pm  \left( \frac{m_\pi}{m_K},  \frac{m_\eta}{m_K} \right) ~.
\end{equation}
While we do not  have any control on the counterterms, we have 
computed the contributions to $f_\pm$  from the $\pi \pi$ ,  $K \bar K$,  $\eta \eta$ loops and find 
$f_\pm  \sim O(1)$,  implying  50\% corrections to  Eq.\ \eqref{eq:epsK0}.

Short-distance corrections to $\ep_K$ arise from $\Delta S = 2$ box diagrams with insertions of 
the right-handed couplings $\xi$. In the case of $\xi_{ud}$ and $\xi_{us}$, diagrams that are linear in $\xi_{ij}$ are necessarily proportional to one power of the 
mass of the internal up quark,
and, using the equations of motion, to the masses of the external $d$ or $s$ quarks. 
This leads to a suppression of $m_u m_s/m_W^2$ that 
makes $O(\xi)$  contributions to dimension-six   $\Delta S = 2$ operators~\cite{Buras:2000if} irrelevant for our analysis.

In summary, for our estimate of right-handed CC contributions to $\ep_K$ we use 
\begin{equation}
\delta \ep_K =  \frac{e^{i \frac{\pi}{4}}}{\sqrt{2}} \left( \frac{\textrm{Im}\,M_{12}(\xi)}{\Delta m_K} + \frac{\textrm{Im}\, A_0(\xi)}{\textrm{Re}\, A_0}  \right),
\label{eq:epkf}
\end{equation}
together with  Eqs. \eqref{A0A2relation} and \eqref{eq:epsK0}, assigning a  50\% uncertainty to this result.

\section{Contributions to  hadronic and nuclear EDMs}\label{sec:EDMs}

\begin{table}[t]
\begin{center}\small
\begin{tabular}{||c||ccccc||}
\hline
 & $d_n$& $d_{\mathrm{Hg}}$ & $d_{\mathrm{Xe}}$ & $d_{\mathrm{Ra}}$ & $d_{p,D}$  \\
\hline
\rule{0pt}{3ex}
current limit  &$ 3.0 \cdot 10^{-13}$  & $6.2 \cdot 10^{-17}$    & $5.5 \cdot 10^{-14}$   & $1.2\cdot 10^{-10}$ & x \\
expected limit &$ 1.0 \cdot 10^{-15} $ & $6.2 \cdot 10^{-17}$   & $5.0 \cdot 10^{-16}$ & $1.0 \cdot 10^{-14}$ &$ 1.0 \cdot 10^{-16}$ \\
\hline
\end{tabular}
\end{center}
\caption{\small Current limits on the neutron \cite{Baker:2006ts,Afach:2015sja}, mercury \cite{Griffith:2009zz,Graner:2016ses}, xenon \cite{PhysRevLett.86.22} and radium \cite{Bishof:2016uqx,Parker:2015yka}
EDMs in units of $e$ fm ($90\%$ confidence level). We also show future sensitivities~\cite{Kumar:2013qya,Chupp:2014gka,Eversmann:2015jnk}.}
\label{tab:EDMexps}  
\end{table}

The operators in  Eq.~\eqref{eq:4q4} also contribute to  hadronic and nuclear EDMs,  
whose  current limits and expected  sensitivities are summarized  in Table \ref{tab:EDMexps}.  
The calculation of  the nuclear EDMs in terms of the operators in Eq.~\eqref{eq:4q4} involves  first matching to an extension of chiral effective field theory (EFT) that contains CPV hadronic interactions 
\cite{deVries:2012ab,Bsaisou:2014oka}.
The chiral  power counting predicts that for the four-quark operators in Eq.~\eqref{eq:4q4}   CPV moments of nuclei are dominated by long-range pion-exchange between nucleons \cite{deVries:2012ab,Bsaisou:2014oka}\footnote{It should be stressed that chiral power counting has not been tested for systems as large as $^{199}$Hg or $^{225}$Ra.}. The leading CPV interactions are 
 \begin{equation}\label{piN}
\mathcal L_{\pi N}^{\rm CPV} = - \frac{\bar g_0}{2 F_\pi} \bar N \boldtau \cdot \boldpi N -  \frac{\bar g_1}{2 F_\pi} \pi_0 \bar N  N\ ,
\end{equation}
where $N=(p\,n)^T$ is the nucleon isospin doublet, $\boldpi$ the pion triplet, and $\bar g_{0,1}$ two LECs that are determined below.  Nuclear calculations,   
within large uncertainties,  predict 
\cite{deJesus:2005nb,Dobaczewski:2005hz,Ban:2010ea,Dzuba:2009kn,Engel:2013lsa,deVries2011b,Bsaisou:2014oka, Singh:2014jca, Singh:2015aba,Yamanaka:2016umw}\footnote{Here we only have given the leading contributions. For $d_D$ there appears an additional contribution $(0.94\pm 0.01)(d_n + d_p)$ \cite{Yamanaka:2015qfa} from the neutron, $d_n$, and proton, $d_p$, EDMs. Similary, $d_{\mathrm{Hg}}$ obtains a contribution $ -(2.8\pm 0.6)\Ex{-4}\left[(1.9\pm0.1)d_n +(0.20\pm 0.06)d_p\right]$ \cite{Dmitriev:2003sc}. These corrections are formally higher order \cite{deVries:2012ab}. The nucleon-EDM contributions to $d_{\mathrm{Xe}}$ and $d_{\mathrm{Ra}}$ have, as far as we know, not been calculated.}
\bea\label{eq:NuclEDM}
d_D &=&- (0.18 \pm 0.02) \frac{\bar g_1}{2F_\pi}\, e \, {\rm fm}\ ,\nn\\ 
d_{\mathrm{Hg}} &=& (2.8\pm 0.6)\Ex{-4}\cdot \bigg(0.13^{+0.5}_{-0.07}\, \frac{\bar g_0}{2F_\pi} + 0.25^{+0.89}_{-0.63}\,\frac{\bar g_1}{2F_\pi}\bigg)e\, {\rm fm}\ ,  \nn \\
d_{\mathrm{Xe}} &=& (0.33\pm 0.05)\Ex{-4}\cdot\left( 0.10_{-0.037}^{+0.53}\, \frac{\bar g_0}{2F_\pi} + 0.076_{-0.038}^{+0.55}\, \frac{\bar g_1}{2F_\pi}\right)e\, {\rm fm}\ , \nn\\
d_{\mathrm{Ra}} &=& (7.7\pm 0.8)\Ex{-4}\cdot\left(-19^{+6.4}_{-57}\,\frac{\bar g_0}{2F_\pi} + 76^{+227}_{-25}\,\frac{\bar g_1}{2F_\pi}\right)e\, {\rm fm}\ ~, 
\eea
where the small prefactors in front of the brackets for the atomic EDMs are the Schiff screening factors~\cite{Schiff:1963zz}.

The other relevant EDMs are those of the neutron and proton.  
For the CPV four-quark operators in Eq.~\eqref{eq:4q4}, the nucleon EDMs are expected to be smaller than the deuteron EDM because 
they do not receive contributions from CPV pion exchange between nucleons. Instead, the CPV pion-nucleon 
interactions in Eq.~\eqref{piN} contribute only at the loop level.

In order to assess the impact of EDMs on the phenomenology of right-handed currents we need 
to determine  $\bar g_{0,1}$ in terms of ${\rm Im} \, \xi_{ud,us}$. The pion-nucleon couplings receive two leading-order contributions in ChPT \cite{Pospelov_qCEDM, Pospelov_piN, deVries:2012ab}. The first one is a ``direct'' contribution, involving the matrix elements 
of the operators in Eq.\ \eqref{eq:4q4} between two nucleons and one pion state, which, currently, we have no control over. 
The second contribution is induced after vacuum alignment through the BSM contributions to $\chi$ in Eq.~\eqref{chi}. As we will argue below, we expect the second piece to provide the dominant contribution to nuclear and diamagnetic EDMs. Since this piece can be related to the $K \rightarrow \pi \pi$ amplitude, it provides a direct connection between $\ep^\prime/\ep$ and EDMs. \\

\noindent {\bf  CP-violating pion-nucleon couplings:}
The leading CPV pNG-baryon couplings arise from the baryonic Lagrangian
\begin{eqnarray}\label{eq:L2}
\mathcal L_{\pi N} &=&   b_0 \textrm{Tr} \left(\bar B_{} B\right) \textrm{Tr} \chi_+  +  b_D \textrm{Tr}\left(\bar B \{ \chi_+, B \}  \right) + b_F \textrm{Tr}\left(\bar B [ \chi_+, B ]  \right) 
+ \mathcal L_{LR} \, ,
\end{eqnarray}
where $B$ denotes the octet baryon field
\begin{equation}\label{eq:3.0}
B = \left( \begin{array}{c c c}
\frac{1}{\sqrt{2}}\Sigma^0 + \frac{1}{\sqrt{6}}\Lambda & \Sigma^+ 							& p \\
\Sigma^-					       & -\frac{1}{\sqrt{2}}\Sigma^0 + \frac{1}{\sqrt{6}}\Lambda 	& n \\
\Xi^-						       & \Xi^0								& -\frac{2}{\sqrt{6}} \Lambda
\end{array} \right)\, .
\end{equation}
The right-handed currents enter through $\chi_+ = u^{\dagger} \chi u^{\dagger} + u \chi^{\dagger} u$  (see Eq.~\eqref{chi}) and  $\vL_{LR}$, which  includes the above-mentioned ``direct" contributions to CP-even baryon masses (from the real parts of the four-quark operators) and CPV pNG-baryons vertices (from the imaginary parts).
$b_{0,D,F}$ induce baryon mass splittings
and,  if the quark mass has a complex component, as dictated by \eqref{chi} and  \eqref{align},
they also induce the CPV pNG-baryon interactions of Eq. \eqref{piN}, with strength 
\begin{eqnarray}\label{relations0}
\bar g_0 &=&  2 (b_D + b_F) F_0^2 \sum_{i=1,2}\mathcal A_{i\, LR} {\rm Im}\, C_{i\, LR}^{us\, us} -8 (b_D + b_F) B m_* (\bar \theta-\bar \theta_{\rm ind})
+\bar g_0|_{\rm direct}  \ , \nn \\
\bar g_1 &=&  2 (2 b_0 + b_D + b_F) F_0^2 \sum_{i=1,2} \mathcal A_{i\, LR} {\rm Im}\, \big(2C_{i\, LR}^{ud\,ud}+C_{i\, LR}^{us\,us}\big) + \bar g_1|_{\rm direct} \, , 
\end{eqnarray}
where we indicated by $\bar g_{0,1} |_{\rm direct}$  the contributions from $\vL_{LR}$. 
In principle, we can now insert values for $b_{0,D,F}$ to obtain estimates for $\bar g_{0,1}$. However, it is possible to improve these relations. First of all, although the direct pieces are unknown at the moment, it is possible to obtain them from LQCD calculations of the baryon mass spectrum induced by the real parts of the four-quark operators. This strategy was suggested for CPV quark chromo-electric dipole moments \cite{AWL,Seng:2016pfd}, but works as well for the operators we consider here. 
Second, 
it is possible to include next-to-leading-order corrections into the relations for $\bar g_{0,1}$ by replacing the combinations of $b_{0,D,F}$ appearing in Eq.~\eqref{relations0} by the strong nucleon mass splitting and nucleon sigma term (see  Refs.~\cite{Seng:2016pfd,AWL,RHC} for  details). Taken together, we obtain
\begin{eqnarray}\label{relationsA}
\bar g_0 &=& - \sum_{i=1,2} \textrm{Im}\, C^{us\, us}_{i\, LR}   \left( \frac{d}{d \textrm{Re}\, C^{us\, us}_{i\, LR}} + \frac{r_i}{4} \frac{d}{d \bar m \varepsilon} \right) \delta m_N + \delta m_N \frac{m_* }{\bar m \varepsilon}(\bar \theta-\bar \theta_{\rm ind})\ , \nn \\
\bar g_1 &=& 2  \sum_{i=1,2}\textrm{Im}\, C^{us\, us}_{i\, LR}  \left( \frac{d}{d \textrm{Re}\, C^{us\, us}_{i\, LR}} - \frac{r_i}{4} \frac{d}{d \bar m } \right)  m_N
+ 4 \sum_{i=1,2}\textrm{Im}\, C^{ud\, ud}_{i\, LR}  \left( \frac{d}{d \textrm{Re}\, C^{ud\, ud}_{i\, LR}} - \frac{r_i}{4} \frac{d}{d \bar m } \right)  m_N\ , 
\nn \\\label{relationsB}
\end{eqnarray}
where $\delta m_N = m_n -m_p$ and $2 m_N = m_n + m_p$ 
and the derivatives  $d (\delta m_N , m_N)/d \textrm{Re}\, C^{us(d)\, us(d)}_{i\, LR}$ 
can be extracted from lattice calculations. The 
tadpole-induced pieces, proportional to $r_i$, depend on known quantities such as the nucleon sigma term $\sigma_N = \bar m (d m_N/d \bar m) = 59.1\pm 3.5$ MeV \cite{Hoferichter:2015dsa} where $\bar m = (m_u+m_d)/2 = 3.37\pm0.08$ MeV \cite{Aoki:2016frl},  
and the nucleon mass induced by the quark mass difference: $(d\delta m_N/d \bar m \varepsilon) \simeq \delta m_N/( \bar m \varepsilon) = (2.49 \pm 0.17 \, \mathrm{MeV})/( \bar m \varepsilon)$ \cite{Borsanyi:2013lga,Borsanyi:2014jba}, where $\varepsilon = (m_d - m_u)/(2\bar m) = 0.37\pm0.03$ \cite{Aoki:2016frl}.

We assume a Peccei-Quinn mechanism such that $\tb$ relaxes to $\tb_{\rm ind}$ and obtain values for $\bar g_{0,1}$ 
\begin{eqnarray}\label{couplings0}
\frac{\bar{g}_0}{2 F_\pi} &=& - (0.16 \pm 0.03\pm 0.08)  \times 10^{-5}\, \textrm{Im} (V^*_{us} \xi_{us})  \,, \\
\label{couplings1}
\frac{\bar{g}_1}{2 F_\pi} &=& - \left( 2.9 \pm 0.33\pm 1.5 \right) \times 10^{-5}\,  \textrm{Im} (V^*_{us} \xi_{us})   -  \left(  5.7 \pm 0.67 \pm 2.9    \right) \times 10^{-5}\,  \textrm{Im} (V^*_{ud} \xi_{ud})\, .
\end{eqnarray}
Here the first error arises from the uncertainties on $\mathcal A_{i\, LR}$, the strong mass splitting and the sigma term, while the second from the unknown direct pieces. We have assigned a $50\%$ uncertainty to the latter which we expect to capture the unknown contributions. In particular, the relatively large values of $\mathcal A_{1,2\, LR}$ (and thus $r_i$) and of the nucleon sigma term cause the indirect contribution to $\bar g_1$ to be enhanced by roughly an order of magnitude over naive-dimensional-analysis estimates of the direct piece \cite{deVries:2012ab}, 
namely $\bar{g}_{0,1}/(2 F_0) \sim  G_F   F_0 \Lambda_\chi  \, {\rm Im} (V_{ij}^* \xi_{ij}) \sim 3 \times 10^{-6} \, {\rm Im} (V_{ij}^* \xi_{ij}) $. We therefore expect Eqs.~\eqref{couplings0} and \eqref{couplings1} to describe the  dominant   contributions to  the CPV couplings. More precise statements require LQCD calculations of the direct contributions in Eq.~\eqref{relationsA}. 
From Eq.~\eqref{couplings0} we see that $\xi_{ud}$ does not contribute to $\bar g_0$ at this order. The exact cancellation is a consequence of the assumed Peccei-Quinn mechanism, but even without this mechanism we would have $\bar g_1 \gg \bar g_0$ \cite{deVries:2012ab, Bsaisou:2014oka}. The same hierarchy is present for $\xi_{us}$ and we conclude that right-handed CCs predominantly lead to isovector CPV nuclear interactions. 
\\

\noindent{\bf The nucleon EDM}\label{sec:nEDM}
For chiral-breaking sources, the CPV pNG-nucleon couplings determine the leading non-analytic contribution to the nucleon EDM. 
We find \cite{Mereghetti:2010kp,Seng:2014pba}
\begin{eqnarray}
d_n &=& \bar d_n  (\mu) + \frac{e g_A \bar g_1}{(4\pi F_\pi)^2} \left(  \frac{\bar g_0}{\bar g_1} \left( \log \frac{m^2_\pi}{\mu^2} - \frac{\pi m_\pi}{2 m_N} \right)  + \frac{1}{4 } \left( \kappa_1 - \kappa_0\right) \frac{m^2_\pi}{m_N^2} \log \frac{m^2_\pi}{\mu^2}  \right) \ ,
\label{eq:dn}
\\
d_p &=& \bar d_p (\mu) 
- \frac{e g_A \bar g_1}{(4\pi F_\pi)^2} \Bigg( \frac{\bar g_0}{\bar g_1} \left( \log \frac{m^2_\pi}{\mu^2} - \frac{2 \pi m_\pi}{m_N} \right)  
-\frac{1}{4 } \left(  \frac{2 \pi m_\pi}{m_N} + \left( \frac{5}{2} + \kappa_1 + \kappa_0\right) \frac{m^2_\pi}{m_N^2} \log \frac{m^2_\pi}{\mu^2}  \right)  \Bigg) 
\ ,\nn\\
\label{eq:dp}
\end{eqnarray}
where $g_A \simeq 1.27$ is the nucleon axial charge, 
and $\kappa_1 = 3.7$ and $\kappa_0 = -0.12$ are related to the nucleon magnetic moments. 
Since $\bar g_1$ is the largest coupling, we included also NLO corrections and  large N${}^2$LO corrections proportional to the nucleon magnetic moment \cite{Seng:2014pba}.   $\bar d_{n,p} (\mu) $  are two counterterms.   
In what follows we set the scale $\mu = m_N$ and use as central values for our estimates 
$\bar d_{n,p}   (\mu = m_N) = 0$.  This leads to 
$d_{n}   = (  1.5   \   \textrm{Im} (V_{ud}^* \xi_{ud} )  +   3.1 \  \textrm{Im} (V_{us}^* \xi_{us} )  )    \times 10^{-7}  \,e \, {\rm fm}   $,  
$d_{p} = - (  1.8  \   \textrm{Im} (V_{ud}^* \xi_{ud} )  + 3.3  \  \textrm{Im} (V_{us}^* \xi_{us} )  )    \times 10^{-7}   \, e \, {\rm fm}  $. 
These values confirm the
 expectation that the EDMs of nuclei and diamagnetic atoms are dominated by pion-exchange contributions \cite{deVries:2012ab}.

The finite parts of the counterterms  cannot be determined by symmetry considerations alone, and constitute an additional source of uncertainties on the nucleon EDM. 
They  can be fixed by matching to  a full non-perturbative calculation,  that could be provided by LQCD in the future. 
For the moment, we  set   $\bar d_{n,p} (\mu)  = \pm  \Delta_{n,p}$ where  
$\Delta_{n} = (  0.5   \   \textrm{Im} (V_{ud}^* \xi_{ud} )  +  1.0   \  \textrm{Im} (V_{us}^* \xi_{us} )  )    \times 10^{-7}   \, e \, {\rm fm}  $ and
$\Delta_{p} = (  0.9    \   \textrm{Im} (V_{ud}^* \xi_{ud} )  - 0.3   \  \textrm{Im} (V_{us}^* \xi_{us} )  )    \times 10^{-7}  \, e \, {\rm fm}  $,   are the variation 
of the loop contributions  in Eqs.\ \eqref{eq:dn} and \eqref{eq:dp}   when we change the renormalization scale from $m_K$ to $m_N$.  
By naive dimension analysis one would obtain   $\bar d_{n,p} = \mathcal O(  G_F  F_0^2  \textrm{Im}V_{ij}^* \xi_{ij} /\Lambda_\chi) \sim 0.3  \,  \textrm{Im} (V_{ij}^* \xi_{ij} )\, \times 10^{-7}\, e \, {\rm fm} $ \cite{Seng:2014pba}, 
well within the range implied by the chiral scale variation. 
Finally,  we have also estimated the size of the contributions to $\bar{d}_{n,p}$ induced by strange particles in the loop contributing to $d_{n,p}$ in $SU(3)$ ChPT \cite{ottnad}, 
finding them to be comparable to  $\Delta_{n,p}$~\cite{RHC}.

We thus  conclude that a conservative assessment of the uncertainties in $d_{n,p}$ is obtained by 
varying the couplings $\bar{g}_{0,1}$ according to  (\ref{couplings0}) and \eqref{couplings1} 
and independently varying the chiral loop scale between $m_K$ and $m_N$,  while setting ${\bar d}_{n,p} (\mu = m_N) = 0$.

\begin{figure}
\center
\includegraphics[width=7.5cm]{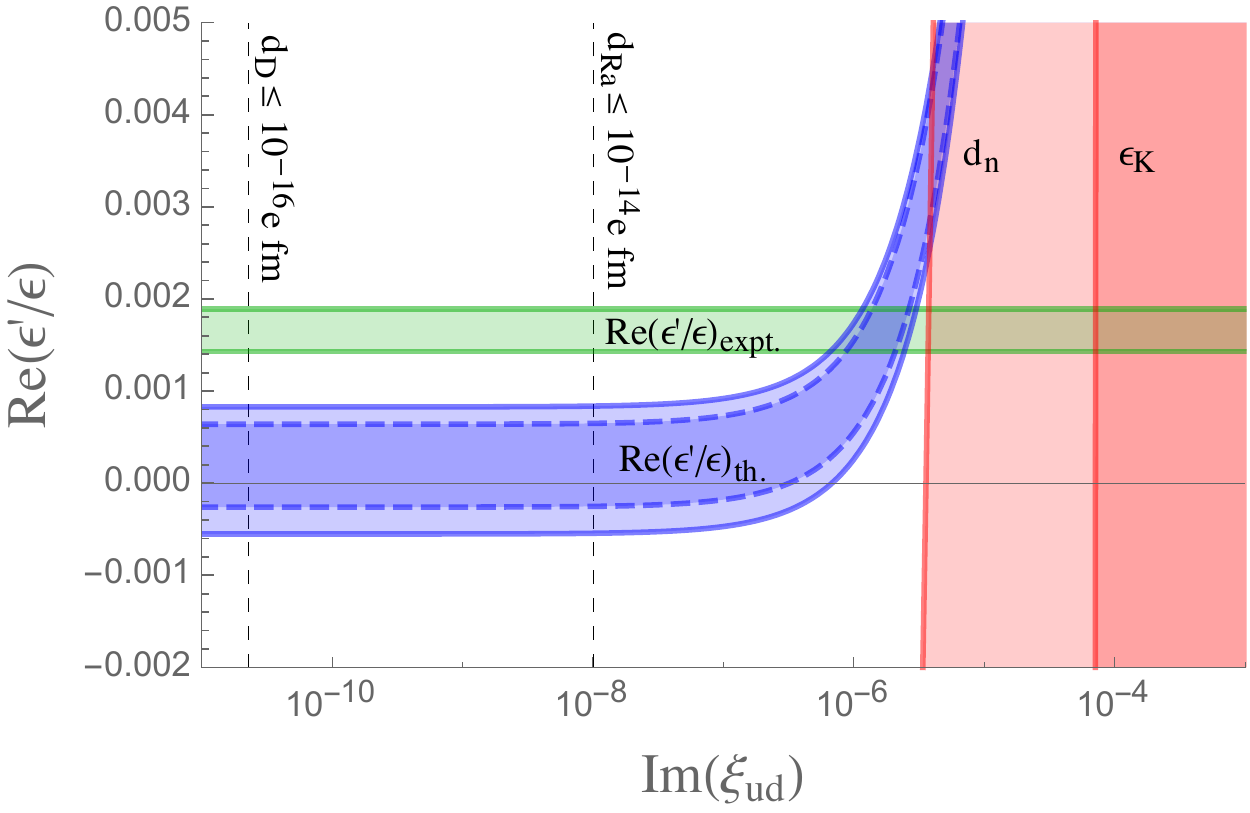}
\quad
\includegraphics[width=7.5cm]{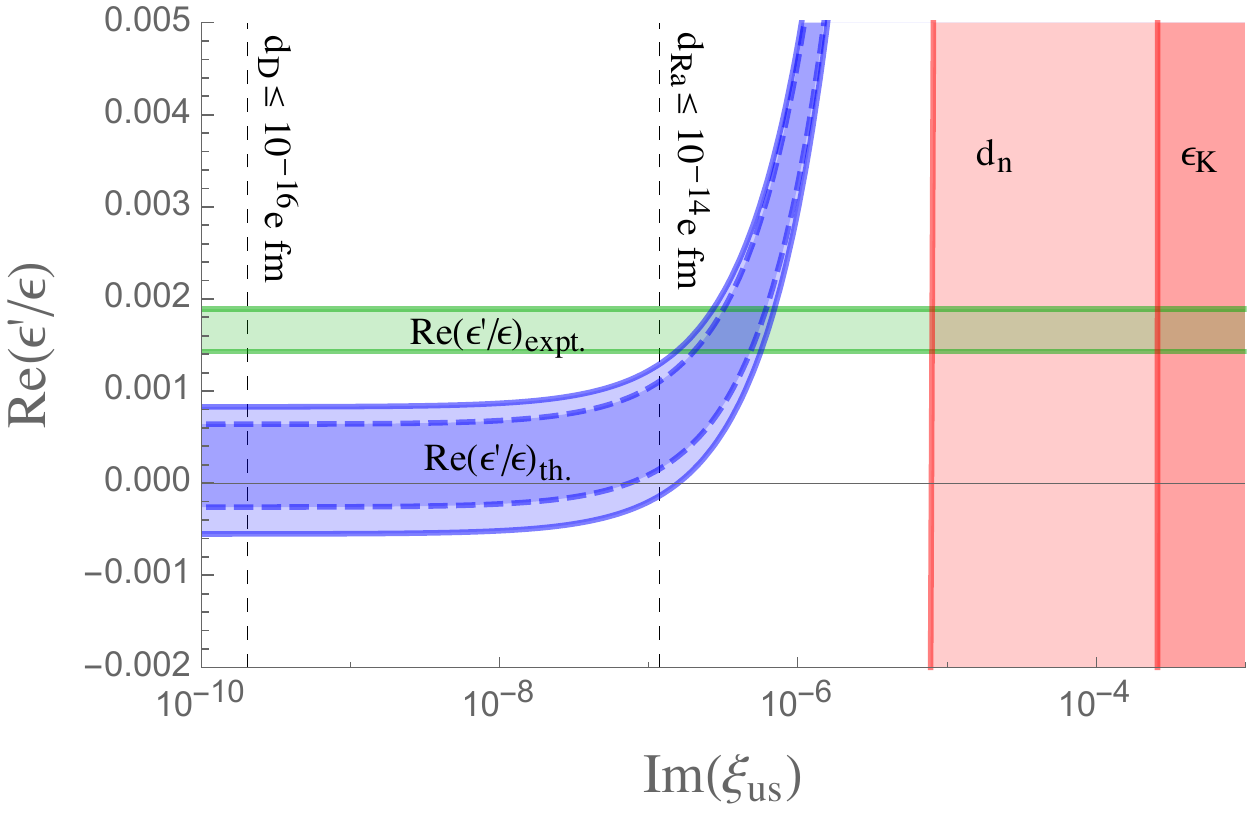}
\vspace{1.0cm}
\includegraphics[width=7.5cm]{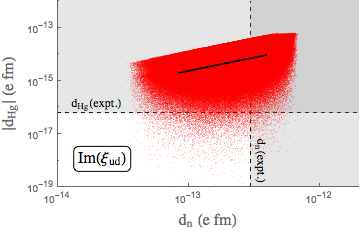}
\quad
\includegraphics[width=7.5cm]{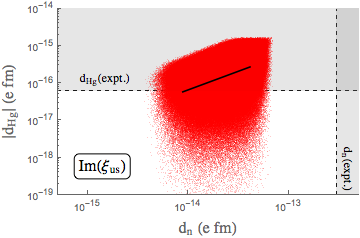}
\includegraphics[width=7.5cm]{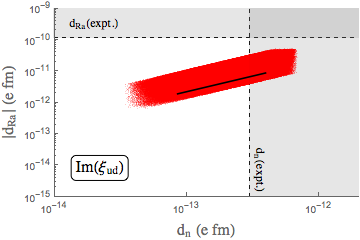}
\quad
\includegraphics[width=7.5cm]{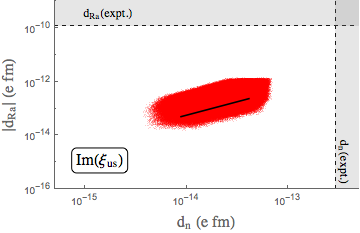}

\caption{\small 
The top-left (-right) panel shows the value of ${\rm Re}\,\ep'/\ep$ as a function of Im$\,\xi_{ud}$ (Im$\,\xi_{us}$). 
The solid blue bands indicate the theoretical value of  ${\rm Re}\,\ep'/\ep$ using $(\ep' / \ep)_{\rm SM} =  (1.4 \pm 6.9) \times 10^{-4}$ based on Ref.~\cite{Bai:2015nea}, while the dashed blue lines apply $(\ep' / \ep)_{\rm SM} =  (1.8 \pm 4.5) \times 10^{-4}$ from Ref.~\cite{Buras:2015yba}, and the experimental value is shown in green (all at $1\, \sigma$). The vertical lines indicate the current/future sensitivities of $\epsilon_K$ and  $d_{n,D,Ra}$ experiments, derived using the R-fit procedure.
The middle-left (-right) panel shows the sizes of $d_{\rm Hg}$ and $d_n$, assuming a value for Im$\,\xi_{ud}$ (Im$\,\xi_{us}$) that solves the $\ep'/\ep$ discrepancy. The red points are generated by taking random values of the nuclear and hadronic matrix elements within their allowed ranges. The black lines result from taking the central values of these matrix elements. The lower two panels show the analogous correlation in the $d_{\rm Ra}-d_n$ plane.
}\label{Fig1}
\end{figure}

\section{$\ep'/\ep$ versus EDMs}\label{sec:analysis}

\noindent{\bf Region of interest:} 
We first investigate what ranges of $\xi_{ud}$ and/or $\xi_{us}$ would align the theoretical predictions of $\ep'/\ep$  with the experimental measurements. 
By combining the results in Sec.~\ref{LowEnergy} and \ref{sec:eps'} we draw the theoretical prediction of $\mathrm{Re}\,\ep'/\ep$ as a function of $\mathrm {Im}\,\xi_{ud}$ ($\mathrm {Im}\,\xi_{us}$) in the upper-left (-right) panel of Fig.~\ref{Fig1}.
Here we assume that only one right-handed coupling is active at a time. The solid blue lines apply the LQCD-based  prediction,
 $(\ep' / \ep)_{\rm SM} =  (1.4 \pm 6.9) \times 10^{-4}$~\cite{Bai:2015nea}, whereas the dashed lines apply the 
  value $(\ep' / \ep)_{\rm SM} =  (1.8 \pm 4.5) \times 10^{-4}$ from Ref.~\cite{Buras:2015yba}.  
From these figures we read off that  $\ep'$ tension would be resolved for couplings in the ranges
\begin{equation}\label{solution}
\mathrm {Im} \, \xi_{ud} \in [0.7,\, 3]\cdot 10^{-6}\ , \qquad \mathrm {Im} \, \xi_{us} \in [1,\, 7] \cdot 10^{-7}\ .
\end{equation}
We can estimate the scale $\Lambda$ where right-handed currents of this size would originate by writing $\mathrm {Im} \, \xi_{ud,us} \sim (v^2/\Lambda^2) \sin \phi_{ud,us}$ in terms of CPV phases $ \sin \phi_{ud,us}$. Assuming these phases to be $ \sin \phi_{ud,us} = \mathcal O(1)$, we obtain  $\Lambda \simeq \{100,\, 300\}$ TeV, a rather high scale. In the context of left-right symmetric models this scale would roughly correspond to masses of $W_R$ bosons. Of course, the right-handed scale can be lowered if the phases are taken to be small. 

\noindent {\bf Constraints from $\ep_K$ and other experiments:} 
$\ep_K$ probes values   of  $\mathrm{Im}\, \xi_{ud,us}$ that are two to three orders of magnitude away from the region of interest \eqref{solution}, as represented by the vertical lines in the upper  panels of Fig.~\ref{Fig1}. 
In obtaining this constraint we use the SM input for $\ep_K$ from Refs.~\cite{Buchalla:1995vs,Buras:2013ooa,Aoki:2016frl,Olive:2016xmw},
which is affected by $\mathcal O(20\%)$ theoretical uncertainties.   
In order to rule out the region of interest,  theoretical/parametric uncertainties on $\ep_K$ need to be reduced below the percent level, an extremely challenging goal.   
In a forthcoming paper \cite{RHC} we perform a global study of right-handed currents including a wide range of experiments. We find that LHC searches probe $|\xi_{ud,us}|$ at the percent level and are thus orders of magnitude away from the above identified region of interest. Leptonic and semi-leptonic pion and kaon decays 
as well as  $\beta$-decays  probe $\mathrm{Re}\,\xi_{ud,us}$ at the $10^{-3,-4}$ level~\cite{Buras:2010pz}. 
Measurements of the triple correlation $\langle \vec J \rangle\cdot (\vec p_e \times \vec p_\nu)$  (the so-called $D$ coefficient)  in  $\beta$-decays
are sensitive to $\mathrm{Im}\,\xi_{ud}$.   For the neutron one has~\cite{Ng:2011ui,Vos:2015eba} 
\begin{eqnarray}
D_n = \frac{4 g_A}{1+3g_A\sq}\, \mathrm{Im}\,\frac{\xi_{ud}}{V_{ud}}\simeq 0.87\, \mathrm{Im}\,\frac{\xi_{ud}}{V_{ud}}\,, 
\end{eqnarray}
which combined with the experimental input  $D= (-1\pm 2.1)\times 10^{-4}$  \cite{Mumm:2011nd} 
results in   $\mathrm{Im}\, \xi_{ud} = (-1.1\pm 4.0)\cdot 10^{-4}$, 
several orders of magnitude away from  the region of interest \eqref{solution}.

\noindent {\bf EDM constraints:}
As discussed in Sec.~\ref{sec:EDMs}, nuclear and diamagnetic EDMs are very sensitive to CPV right-handed currents. The most precise EDM measurement, $d_{\mathrm{Hg}}$, suffers from large nuclear uncertainties, see Eq.~\eqref{couplings0}, in addition to significant hadronic uncertainties, see Eq.~\eqref{eq:NuclEDM}. To handle these uncertainties we apply the  Range-fit (R-fit) procedure defined in Ref. \cite{Charles:2004jd}. 
This strategy provides the most conservative constraints   as it allows for cancellations between different contributions. 
Unfortunately, the theoretical uncertainties of $d_{\mathrm{Hg}}$ are so large that within the R-fit strategy, it is always possible to cancel any constraint on $\mathrm{Im}\,\xi_{ud,us}$. At the moment, $d_{\mathrm{Xe,\,Ra,}}$ are not  sensitive enough and the best constraint arises from $d_n$, depicted by the vertical red lines in the top panels of Fig.~\ref{Fig1}. The $d_n$ limit is consistent with values of $\mathrm{Im}\,\xi_{ud,us}$ that explain $\ep'/\ep$. However, an experimental improvement of one (two) order(s) of magnitude would be sufficient to probe the required value of $\mathrm{Im}\,\xi_{ud}$ ($\mathrm{Im}\,\xi_{us}$). Because the nuclear theory is, respectively, much and somewhat better for the $d_D$ and $d_{\mathrm{Ra}}$, the projected sensitivities of these experiments will also 
be sufficient to rule out a right-handed explanation of $\ep^\prime/\ep$, as illustrated by the vertical dashed black lines.
  
The above analysis is  very conservative  as it removes the impact of the $d_{\mathrm{Hg}}$ limit by marginalizing over the matrix element uncertainties.  In order to graphically show the effect of matrix-element uncertainties and to illustrate the impact of $d_{\mathrm{Hg}}$ we present in the middle panel of Fig.~\ref{Fig1} a scatter plot  of the sizes of $d_n$ and $d_{\mathrm{Hg}}$. Each point is obtained by using a value of $\mathrm{Im}\,\xi_{ud}$ (left panel) or $\mathrm{Im}\,\xi_{us}$ (right panel) that resolves the $\ep'/\ep$ discrepancy. However, for each point we took a random value of the nuclear and hadronic matrix elements within the allowed ranges. The black solid lines indicate the size of $d_n$ and $d_{\mathrm{Hg}}$ when central values of the hadronic and nuclear matrix elements are applied, while scanning $\mathrm{Im}\,\xi_{ud,\, us}$ in the region \eqref{solution}.
   
The middle-left panel illustrates that for most of the matrix elements $d_{\mathrm{Hg}}$ rules out the value of $\mathrm{Im}\,\xi_{ud}$ required to explain $\ep^\prime/\ep$.
This implies that a modest improvement in the nuclear-structure uncertainties could significantly impact the analysis, something also encountered in the analysis of CPV Higgs interactions \cite{Chien:2015xha}. For $\mathrm{Im}\,\xi_{us}$ (middle-right panel) the situation is less severe and for large range of hadronic and nuclear matrix elements the $d_{\mathrm{Hg}}$ limit is consistent with a right-handed solution of $\ep'/\ep$. 
  
Finally, the bottom panels show similar scatter plots, but now for $d_n$ versus $d_{\mathrm{Ra}}$. At the moment, for all matrix elements in the allowed range, the required values of $\mathrm{Im}\,\xi_{ud,us}$ are consistent with the $d_{\mathrm{Ra}}$ limit, while $d_n$ is cutting into some of the matrix elements in case of $\mathrm{Im}\,\xi_{ud}$. A ${}^{225}$Ra EDM limit of $d_{\mathrm{Ra}} \leq 5\times10^{-13} e$ fm ($d_{\mathrm{Ra}} \leq 10^{-14} e$ fm) would probe the entire relevant parameter space of $\xi_{ud}$ ($\xi_{us}$). These values fall within the projected accuracy of the $d_{\mathrm{Ra}}$ program.

\begin{figure}[t]
\center
\includegraphics[width=7.5cm]{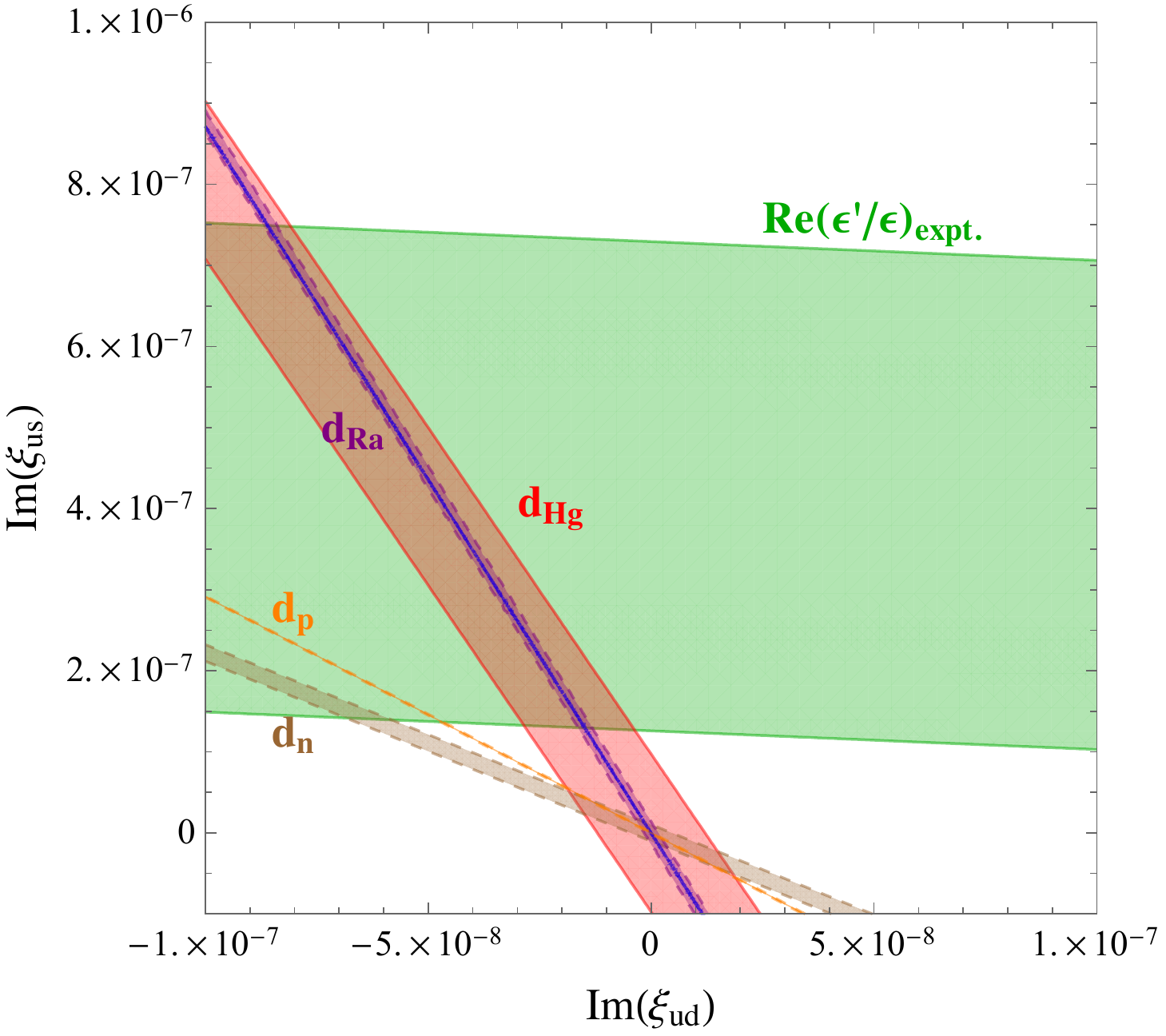}
\caption{\small The figure shows the ${\rm Im}\, \xi_{ud}-{\rm Im}\, \xi_{us}$ plane with the preferred ${\rm Re}\,\ep'/\ep$ region in green. The remaining bands show the current mercury limit and future EDM constraints from $d_n\leq 10^{-15}e$ fm, $d_n\leq 10^{-15}e$ fm, $d_{p,D}\leq 10^{-16}e$ fm, and $d_{\rm Ra}\leq 10^{-14}e$ fm (all at $1$ $\sigma$). The EDM constraints are derived assuming no theoretical uncertainties. The future $d_D$ constraint is depicted by the (unlabeled) thin blue line inside the $d_{\rm Ra}$ band.
}\label{Fig2}
\end{figure}

To thoroughly test a resolution of the $\ep'/\ep$ tension in terms of right-handed CCs, we should consider the possibility that  $\xi_{ud}$ and $\xi_{us}$ are turned on simultaneously. To this end,  Fig.\ \ref{Fig2} shows the preferred ${\rm Re}\,\ep^\prime/\ep$ region (green band) in the  Im$\,\xi_{ud}$-Im$\,\xi_{us}$ plane,  along with  the current $d_{\rm Hg}$ limit and the constraints from $d_{n}$, $d_{p}$, $d_{D}$, and $d_{\rm  Ra}$ at the  future sensitivities indicated in  Table \ref{tab:EDMexps}.
Here we did not consider  theory  uncertainties in the EDM analysis, so  the depicted constraining power of the EDMs, especially that of $d_{\rm Hg}$, is rather optimistic.
Nevertheless, a future $d_{\rm Ra}$  and/or $d_D$  measurement (the purple band and the blue line inside it, respectively) would not fully exclude the parameter space within the preferred ${\rm Re}\,\ep'/\ep$ region. However, complementing such constraints with an improved $d_n$ (brown band) or future $d_p$ (orange band) experiment, would allow us to rule out the solution of the $\ep'/\ep$ discrepancy in terms of $\xi_{ud}$ and $\xi_{us}$.

\section{Conclusions}\label{sec:conclusion}

We have investigated within the context of right-handed charged currents interactions the apparent discrepancy between SM predictions and measurements of direct CP violation in $K\rightarrow \pi \pi$ decays. Although it is too early to tell, if the discrepancy turns out to be real, right-handed charged currents are an attractive solution as they modify $\ep'/\ep$ at tree level. 

We employed chiral perturbation theory in combination with LQCD results to accurately calculate the right-handed contribution to $\ep'/\ep$. 
We identified the required values of right-handed couplings, hinting to a right-handed scale of $\mathcal{O}(10^2\,\mathrm{TeV})$, 
to explain the discrepancy and found that such values are not in conflict with other experiments. 
At the moment, only EDM experiments come close to the required sensitivity. 
By again combining LQCD and chiral techniques we have related CP violation in the kaon sector to hadronic EDMs, 
and  our analysis can be implemented in future investigations of left-right symmetric models. Although current EDM experiments are not in conflict with right-handed interactions that explain the $\ep'/\ep$ discrepancy, next-generation EDM experiments (in particular that of $d_{\mathrm{Ra}}$ and $d_n$) will reach the required sensitivity.  A nonzero EDM signal of the predicted size would provide  tantalizing support for high-scale right-handed charged current interactions.    

\section*{Acknowledgements}
VC and EM  acknowledge support by the US DOE Office of Nuclear Physics and by the LDRD program at Los Alamos National Laboratory.
WD and JdV  acknowledge  support by the Dutch Organization for Scientific Research (NWO) 
through a RUBICON  and VENI grant, respectively. We thank C.-Y.~Seng for discussions on the relation between CPV pion-nucleon couplings and the baryon spectrum, and R. Fleischer for discussions on the impact of $\ep_K$.

\appendix

\bibliographystyle{h-physrev3} 
\bibliography{bibliography}

\begin{thebibliography}{10}

\bibitem{AlaviHarati:2002ye}
KTeV, A.~Alavi-Harati {\em et~al.},
\newblock Phys. Rev. {\bf D67}, 012005 (2003), hep-ex/0208007,
\newblock [Erratum: Phys. Rev.D70,079904(2004)].

\bibitem{Abouzaid:2010ny}
KTeV, E.~Abouzaid {\em et~al.},
\newblock Phys. Rev. {\bf D83}, 092001 (2011), 1011.0127.

\bibitem{Batley:2002gn}
NA48, J.~R. Batley {\em et~al.},
\newblock Phys. Lett. {\bf B544}, 97 (2002), hep-ex/0208009.

\bibitem{Ciuchini:1992tj}
M.~Ciuchini, E.~Franco, G.~Martinelli, and L.~Reina,
\newblock Phys. Lett. B {\bf 301}, 263 (1993), hep-ph/9212203.

\bibitem{Buras:1993dy}
A.~J. Buras, M.~Jamin, and M.~E. Lautenbacher,
\newblock Nucl. Phys. {\bf B408}, 209 (1993), hep-ph/9303284.

\bibitem{Buchalla:1995vs}
G.~Buchalla, A.~J. Buras, and M.~E. Lautenbacher,
\newblock Rev. Mod. Phys. {\bf 68}, 1125 (1996), hep-ph/9512380.

\bibitem{Ciuchini:1995cd}
M.~Ciuchini, E.~Franco, G.~Martinelli, L.~Reina, and L.~Silvestrini,
\newblock Z. Phys. {\bf C68}, 239 (1995), hep-ph/9501265.

\bibitem{Bertolini:1998vd}
S.~Bertolini, M.~Fabbrichesi, and J.~O. Eeg,
\newblock Rev. Mod. Phys. {\bf 72}, 65 (2000), hep-ph/9802405.

\bibitem{Buras:2003zz}
A.~J. Buras and M.~Jamin,
\newblock JHEP {\bf 01}, 048 (2004), hep-ph/0306217.

\bibitem{Pich:2004ee}
A.~Pich,
\newblock {Epsilon-prime/epsilon in the standard model: Theoretical update},
\newblock in {\em {Proceedings, 32nd International Conference on High Energy
  Physics (ICHEP 2004): Beijing, China, August 16-22, 2004}}, 2004,
  hep-ph/0410215.

\bibitem{Cirigliano:2011ny}
V.~Cirigliano, G.~Ecker, H.~Neufeld, A.~Pich, and J.~Portoles,
\newblock Rev. Mod. Phys. {\bf 84}, 399 (2012), 1107.6001.

\bibitem{Bai:2015nea}
RBC, UKQCD, Z.~Bai {\em et~al.},
\newblock Phys. Rev. Lett. {\bf 115}, 212001 (2015), 1505.07863.

\bibitem{Buras:2015yba}
A.~J. Buras, M.~Gorbahn, S.~Jäger, and M.~Jamin,
\newblock JHEP {\bf 11}, 202 (2015), 1507.06345.

\bibitem{Buras:2016fys}
A.~J. Buras and J.-M. Gerard,
\newblock (2016), 1603.05686.

\bibitem{Kitahara:2016nld}
T.~Kitahara, U.~Nierste, and P.~Tremper,
\newblock (2016), 1607.06727.

\bibitem{Bertolini:2000dy}
S.~Bertolini, J.~O. Eeg, and M.~Fabbrichesi,
\newblock Phys. Rev. D {\bf 63}, 056009 (2001), hep-ph/0002234.

\bibitem{Bijnens:2000im}
J.~Bijnens and J.~Prades,
\newblock JHEP {\bf 06}, 035 (2000), hep-ph/0005189.

\bibitem{Pallante:2001he}
E.~Pallante, A.~Pich, and I.~Scimemi,
\newblock Nucl. Phys. B {\bf 617}, 441 (2001), hep-ph/0105011.

\bibitem{Gabbiani:1996hi}
F.~Gabbiani, E.~Gabrielli, A.~Masiero, and L.~Silvestrini,
\newblock Nucl. Phys. {\bf B477}, 321 (1996), hep-ph/9604387.

\bibitem{Buras:2000qz}
A.~J. Buras, P.~Gambino, M.~Gorbahn, S.~Jager, and L.~Silvestrini,
\newblock Nucl. Phys. {\bf B592}, 55 (2001), hep-ph/0007313.

\bibitem{Masiero:1999ub}
A.~Masiero and H.~Murayama,
\newblock Phys. Rev. Lett. {\bf 83}, 907 (1999), hep-ph/9903363.

\bibitem{Babu:1999xf}
K.~S. Babu, B.~Dutta, and R.~N. Mohapatra,
\newblock Phys. Rev. {\bf D61}, 091701 (2000), hep-ph/9905464.

\bibitem{Barbieri:1999ax}
R.~Barbieri, R.~Contino, and A.~Strumia,
\newblock Nucl. Phys. {\bf B578}, 153 (2000), hep-ph/9908255.

\bibitem{Buras:1999da}
A.~J. Buras, G.~Colangelo, G.~Isidori, A.~Romanino, and L.~Silvestrini,
\newblock Nucl. Phys. {\bf B566}, 3 (2000), hep-ph/9908371.

\bibitem{Kagan:1999iq}
A.~L. Kagan and M.~Neubert,
\newblock Phys. Rev. Lett. {\bf 83}, 4929 (1999), hep-ph/9908404.

\bibitem{Bertolini:2012pu}
S.~Bertolini, J.~O. Eeg, A.~Maiezza, and F.~Nesti,
\newblock Phys. Rev. D {\bf 86}, 095013 (2012), 1206.0668,
\newblock [Erratum: Phys. Rev.D93,no.7,079903(2016)].

\bibitem{Bertolini:2014sua}
S.~Bertolini, A.~Maiezza, and F.~Nesti,
\newblock Phys. Rev. {\bf D89}, 095028 (2014), 1403.7112.

\bibitem{Buras:2015jaq}
A.~J. Buras,
\newblock JHEP {\bf 04}, 071 (2016), 1601.00005.

\bibitem{Buras:2015kwd}
A.~J. Buras and F.~De~Fazio,
\newblock JHEP {\bf 03}, 010 (2016), 1512.02869.

\bibitem{Tanimoto:2016yfy}
M.~Tanimoto and K.~Yamamoto,
\newblock (2016), 1603.07960.

\bibitem{Kitahara:2016otd}
T.~Kitahara, U.~Nierste, and P.~Tremper,
\newblock Phys. Rev. Lett. {\bf 117}, 091802 (2016), 1604.07400.

\bibitem{Endo:2016aws}
M.~Endo, S.~Mishima, D.~Ueda, and K.~Yamamoto,
\newblock Phys. Lett. {\bf B762}, 493 (2016), 1608.01444.

\bibitem{Bobeth:2016llm}
C.~Bobeth, A.~J. Buras, A.~Celis, and M.~Jung,
\newblock (2016), 1609.04783.

\bibitem{Zhang:2007da}
Y.~Zhang, H.~An, X.~Ji, and R.~N. Mohapatra,
\newblock Nucl. Phys. B {\bf 802}, 247 (2008), 0712.4218.

\bibitem{An:2009zh}
H.~An, X.~Ji, and F.~Xu,
\newblock JHEP {\bf 1002}, 043 (2010), 0908.2420.

\bibitem{Xu:2009nt}
F.~Xu, H.~An, and X.~Ji,
\newblock JHEP {\bf 1003}, 088 (2010), 0910.2265.

\bibitem{Maiezza:2014ala}
A.~Maiezza and M.~Nemevsek,
\newblock Phys. Rev. {\bf D90}, 095002 (2014), 1407.3678.

\bibitem{Mohapatra:1974hk}
R.~N. Mohapatra and J.~C. Pati,
\newblock Phys. Rev. D {\bf 11}, 566 (1975).

\bibitem{Senjanovic:1975rk}
G.~Senjanovic and R.~N. Mohapatra,
\newblock Phys. Rev. D {\bf 12}, 1502 (1975).

\bibitem{Buchmuller:1982ye}
W.~Buchmuller and D.~Wyler,
\newblock Phys.Lett. {\bf B121}, 321 (1983).

\bibitem{Grzadkowski:2010es}
B.~Grzadkowski, M.~Iskrzynski, M.~Misiak, and J.~Rosiek,
\newblock JHEP {\bf 1010}, 085 (2010), 1008.4884.

\bibitem{Bertolini:2013noa}
S.~Bertolini, A.~Maiezza, and F.~Nesti,
\newblock Phys. Rev. D {\bf 88}, 034014 (2013), 1305.5739.

\bibitem{Chen:2008kt}
P.~Chen, H.~Ke, and X.~Ji,
\newblock Phys. Lett. B {\bf 677}, 157 (2009), 0810.2576.

\bibitem{Blum:2012uk}
T.~Blum {\em et~al.},
\newblock Phys. Rev. D {\bf 86}, 074513 (2012), 1206.5142.

\bibitem{Blum:2015ywa}
T.~Blum {\em et~al.},
\newblock Phys. Rev. D {\bf 91}, 074502 (2015), 1502.00263.

\bibitem{Cho:1993zb}
P.~L. Cho and M.~Misiak,
\newblock Phys. Rev. D {\bf 49}, 5894 (1994), hep-ph/9310332.

\bibitem{Hisano:2012cc}
J.~Hisano, K.~Tsumura, and M.~J. Yang,
\newblock Phys.Lett. {\bf B713}, 473 (2012), 1205.2212.

\bibitem{Dekens:2013zca}
W.~Dekens and J.~de~Vries,
\newblock JHEP {\bf 1305}, 149 (2013), 1303.3156.

\bibitem{Dashen:1970et}
R.~F. Dashen,
\newblock Phys. Rev. {\bf D3}, 1879 (1971).

\bibitem{deVries:2012ab}
J.~de~Vries, E.~Mereghetti, R.~G.~E. Timmermans, and U.~van Kolck,
\newblock Annals Phys. {\bf 338}, 50 (2013), 1212.0990.

\bibitem{Agashe:2014kda}
Particle Data Group, K.~A. Olive {\em et~al.},
\newblock Chin. Phys. {\bf C38}, 090001 (2014).

\bibitem{Peccei:1977hh}
R.~D. Peccei and H.~R. Quinn,
\newblock Phys. Rev. Lett. {\bf 38}, 1440 (1977).

\bibitem{Bijnens:1983ye}
J.~Bijnens and M.~B. Wise,
\newblock Phys. Lett. {\bf B137}, 245 (1984).

\bibitem{Buras:2010pz}
A.~J. Buras, K.~Gemmler, and G.~Isidori,
\newblock Nucl. Phys. {\bf B843}, 107 (2011), 1007.1993.

\bibitem{RHC}
S.~Alioli, V.~Cirigliano, W.~Dekens, J.~de~Vries, and E.~Mereghetti,
\newblock in preparation .

\bibitem{Buras:2010pza}
A.~J. Buras, D.~Guadagnoli, and G.~Isidori,
\newblock Phys. Lett. {\bf B688}, 309 (2010), 1002.3612.

\bibitem{Buras:2000if}
A.~J. Buras, M.~Misiak, and J.~Urban,
\newblock Nucl. Phys. {\bf B586}, 397 (2000), hep-ph/0005183.

\bibitem{Baker:2006ts}
C.~A. Baker {\em et~al.},
\newblock Phys. Rev. Lett. {\bf 97}, 131801 (2006), hep-ex/0602020.

\bibitem{Afach:2015sja}
J.~Pendlebury {\em et~al.},
\newblock Phys. Rev. {\bf D92}, 092003 (2015), 1509.04411.

\bibitem{Griffith:2009zz}
W.~C. Griffith {\em et~al.},
\newblock Phys. Rev. Lett. {\bf 102}, 101601 (2009).

\bibitem{Graner:2016ses}
B.~Graner, Y.~Chen, E.~G. Lindahl, and B.~R. Heckel,
\newblock Phys. Rev. Lett. {\bf 116}, 161601 (2016), 1601.04339.

\bibitem{PhysRevLett.86.22}
M.~A. Rosenberry and T.~E. Chupp,
\newblock Phys. Rev. Lett. {\bf 86}, 22 (2001).

\bibitem{Bishof:2016uqx}
M.~Bishof {\em et~al.},
\newblock Phys. Rev. {\bf C94}, 025501 (2016), 1606.04931.

\bibitem{Parker:2015yka}
R.~Parker {\em et~al.},
\newblock Phys. Rev. Lett. {\bf 114}, 233002 (2015), 1504.07477.

\bibitem{Kumar:2013qya}
K.~Kumar, Z.-T. Lu, and M.~J. Ramsey-Musolf,
\newblock {Working Group Report: Nucleons, Nuclei, and Atoms},
\newblock in {\em {Community Summer Study 2013: Snowmass on the Mississippi
  (CSS2013) Minneapolis, MN, USA, July 29-August 6, 2013}}, 2013, 1312.5416.

\bibitem{Chupp:2014gka}
T.~Chupp and M.~Ramsey-Musolf,
\newblock Phys. Rev. {\bf C91}, 035502 (2015), 1407.1064.

\bibitem{Eversmann:2015jnk}
JEDI, D.~Eversmann {\em et~al.},
\newblock Phys. Rev. Lett. {\bf 115}, 094801 (2015), 1504.00635.

\bibitem{Bsaisou:2014oka}
J.~Bsaisou, U.-G. Mei{\ss}ner, A.~Nogga, and A.~Wirzba,
\newblock Annals Phys. {\bf 359}, 317 (2015), 1412.5471.

\bibitem{deJesus:2005nb}
J.~H. de~Jesus and J.~Engel,
\newblock Phys. Rev. {\bf C72}, 045503 (2005), nucl-th/0507031.

\bibitem{Dobaczewski:2005hz}
J.~Dobaczewski and J.~Engel,
\newblock Phys. Rev. Lett. {\bf 94}, 232502 (2005), nucl-th/0503057.

\bibitem{Ban:2010ea}
S.~Ban, J.~Dobaczewski, J.~Engel, and A.~Shukla,
\newblock Phys. Rev. {\bf C82}, 015501 (2010), 1003.2598.

\bibitem{Dzuba:2009kn}
V.~A. Dzuba, V.~V. Flambaum, and S.~G. Porsev,
\newblock Phys. Rev. A {\bf 80}, 032120 (2009), 0906.5437.

\bibitem{Engel:2013lsa}
J.~Engel, M.~J. Ramsey-Musolf, and U.~van Kolck,
\newblock Prog. Part. Nucl. Phys. {\bf 71}, 21 (2013), 1303.2371.

\bibitem{deVries2011b}
J.~de~Vries {\em et~al.},
\newblock Phys. Rev. C {\bf 84}, 065501 (2011), 1109.3604.

\bibitem{Singh:2014jca}
Y.~Singh and B.~K. Sahoo,
\newblock Phys. Rev. {\bf A91}, 030501 (2015), 1408.4337.

\bibitem{Singh:2015aba}
Y.~Singh and B.~K. Sahoo,
\newblock Phys. Rev. {\bf A92}, 022502 (2015), 1504.00269.

\bibitem{Yamanaka:2016umw}
N.~Yamanaka,
\newblock (2016), 1609.04759.

\bibitem{Yamanaka:2015qfa}
N.~Yamanaka and E.~Hiyama,
\newblock Phys. Rev. {\bf C91}, 054005 (2015), 1503.04446.

\bibitem{Dmitriev:2003sc}
V.~F. Dmitriev and R.~A. Sen'kov,
\newblock Phys. Rev. Lett. {\bf 91}, 212303 (2003), nucl-th/0306050.

\bibitem{Schiff:1963zz}
L.~Schiff,
\newblock Phys.Rev. {\bf 132}, 2194 (1963).

\bibitem{Pospelov_qCEDM}
M.~Pospelov and A.~Ritz,
\newblock Phys. Rev. D {\bf 63}, 073015 (2001), hep-ph/0010037.

\bibitem{Pospelov_piN}
M.~Pospelov,
\newblock Phys. Lett. B {\bf 530}, 123 (2002), hep-ph/0109044.

\bibitem{AWL}
J.~de~Vries, E.~Mereghetti, C.-Y. Seng, and A.~Walker-Loud,
\newblock (2016), 1612.01567.

\bibitem{Seng:2016pfd}
C.-Y. Seng and M.~Ramsey-Musolf,
\newblock (2016), 1611.08063.

\bibitem{Hoferichter:2015dsa}
M.~Hoferichter, J.~Ruiz~de Elvira, B.~Kubis, and U.-G. Mei$\ss$ner,
\newblock Phys. Rev. Lett. {\bf 115}, 092301 (2015), 1506.04142.

\bibitem{Aoki:2016frl}
S.~Aoki {\em et~al.},
\newblock (2016), 1607.00299.

\bibitem{Borsanyi:2013lga}
S.~Borsanyi {\em et~al.},
\newblock Phys.Rev.Lett. {\bf 111}, 252001 (2013), 1306.2287.

\bibitem{Borsanyi:2014jba}
S.~Borsanyi {\em et~al.},
\newblock Science {\bf 347}, 1452 (2015), 1406.4088.

\bibitem{Mereghetti:2010kp}
E.~Mereghetti, J.~de~Vries, W.~H. Hockings, C.~M. Maekawa, and U.~van Kolck,
\newblock Phys. Lett. {\bf B696}, 97 (2011), 1010.4078.

\bibitem{Seng:2014pba}
C.-Y. Seng, J.~de~Vries, E.~Mereghetti, H.~H. Patel, and M.~Ramsey-Musolf,
\newblock Phys. Lett. {\bf B736}, 147 (2014), 1401.5366.

\bibitem{ottnad}
K.~Ottnad, B.~Kubis, U.-G. Mei{\ss}ner, and F.-K. Guo,
\newblock Phys. Lett. B {\bf 687}, 42 (2010), 0911.3981.

\bibitem{Buras:2013ooa}
A.~J. Buras and J.~Girrbach,
\newblock Rept. Prog. Phys. {\bf 77}, 086201 (2014), 1306.3775.

\bibitem{Olive:2016xmw}
Particle Data Group, C.~Patrignani {\em et~al.},
\newblock Chin. Phys. {\bf C40}, 100001 (2016).

\bibitem{Ng:2011ui}
J.~Ng and S.~Tulin,
\newblock Phys. Rev. D {\bf 85}, 033001 (2012), 1111.0649.

\bibitem{Vos:2015eba}
K.~K. Vos, H.~W. Wilschut, and R.~G.~E. Timmermans,
\newblock Rev. Mod. Phys. {\bf 87}, 1483 (2015), 1509.04007.

\bibitem{Mumm:2011nd}
H.~P. Mumm {\em et~al.},
\newblock Phys. Rev. Lett. {\bf 107}, 102301 (2011), 1104.2778.

\bibitem{Charles:2004jd}
CKMfitter Group, J.~Charles {\em et~al.},
\newblock Eur. Phys. J. {\bf C41}, 1 (2005), hep-ph/0406184.

\bibitem{Chien:2015xha}
Y.~T. Chien, V.~Cirigliano, W.~Dekens, J.~de~Vries, and E.~Mereghetti,
\newblock JHEP {\bf 02}, 011 (2016), 1510.00725,
\newblock [JHEP02,011(2016)].

\end{thebibliography}

\end{document}